\documentclass[sigconf,natbib=false,anonymous=false]{acmart}

\usepackage{balance}

\AtBeginDocument{%
  }

\copyrightyear{2024}
\acmYear{2024}
\setcopyright{acmlicensed}\acmConference[SIGIR '24]{Proceedings of the 47th International ACM SIGIR Conference on Research and Development in Information Retrieval}{July 14--18, 2024}{Washington, DC, USA}
\acmBooktitle{Proceedings of the 47th International ACM SIGIR Conference on Research and Development in Information Retrieval (SIGIR '24), July 14--18, 2024, Washington, DC, USA}
\acmDOI{10.1145/3626772.3657924}
\acmISBN{979-8-4007-0431-4/24/07}

\RequirePackage[
  datamodel=acmdatamodel,
  style=acmnumeric,
  ]{biblatex}

\begin{document}

\title{Behavior Alignment: A New Perspective of Evaluating LLM-based Conversational Recommender Systems}



\author{Dayu Yang}
\affiliation{%
    \institution{University of Delaware}
    \city{Newark}
    \state{DE}
    \country{USA}}
\email{dayu@udel.edu}

\author{Fumian Chen}

\affiliation{%
  \institution{University of Delaware}
  \city{Newark}
  \state{DE}
  \country{USA}}
\email{fmchen@udel.edu}

\author{Hui Fang}

\affiliation{%
  \institution{ University of Delaware}
  \city{Newark}
  \state{DE}
  \country{USA}}
\email{hfang@udel.edu}

\renewcommand{\shortauthors}{Dayu Yang, Fumian Chen, \& Hui Fang}

\begin{abstract}
Large Language Models (LLMs) have demonstrated great potential in Conversational Recommender Systems (CRS). However, the application of LLMs to CRS has exposed a notable discrepancy in behavior between LLM-based CRS and human recommenders: LLMs often appear inflexible and passive, frequently rushing to complete the recommendation task without sufficient inquiry.
This behavior discrepancy can lead to decreased accuracy in recommendations and lower user satisfaction. 
Despite its importance, existing studies in CRS lack a study about how to measure such behavior discrepancy. 
To fill this gap, we propose {\em Behavior Alignment}, a new evaluation metric to measure how well the recommendation strategies made by a LLM-based CRS are consistent with human recommenders'. 
Our experiment results show that the 
new metric is better aligned with human preferences and can better differentiate \textit{how systems perform} than existing evaluation metrics.  
As Behavior Alignment requires explicit and costly human annotations on the recommendation 
strategies, we also propose a classification-based method 
to implicitly measure the Behavior Alignment based on the responses. The evaluation 
results confirm the robustness of the method.
\end{abstract}




\maketitle

\section{Introduction}

Recent advancements have highlighted the potential of large language models (LLMs) across a range of tasks. Their impressive recommendation performance~\cite{sileo2022zero, hou2023large, zhang2023recommendation} and language generation capability~\cite{wei2022emergent} have attracted the attention of researchers in the conversational recommender system (CRS) community~\cite{friedman2023leveraging, liao2023proactive, zhang2023user, he2023large}. For example, Google Research team~\cite{friedman2023leveraging} recently proposed an LLM-based CRS that achieved huge success on YouTube. LLMs are inherently well-suited to the needs of CRS. Unlike traditional recommender systems, which depend on user profiles and past activities, CRS prioritizes the identification of user preferences through real-time interactions~\cite{sun2018conversational, zhang2023user, friedman2023leveraging}, striving to mimic the interactions of human recommenders~\cite{hayati2020inspired, li2018towards}. These characteristics require strong capabilities in language understanding and generation~\cite{li2018towards,zhou2020improving}, areas where LLMs excel. Consequently, there is an increasing trend of adopting LLMs within the CRS domain.

However, existing studies also found a significant weakness of LLMs when implemented into conversational recommendation~\cite{zhang2023user, liao2023proactive}: the behavior of LLMs lacks proactiveness and adaptivity throughout dialogues, therefore leads to the shortage of information for understanding the users' preference~\cite{liao2023proactive}. 

\begin{figure}
    \centering
    \includegraphics[scale=0.28]{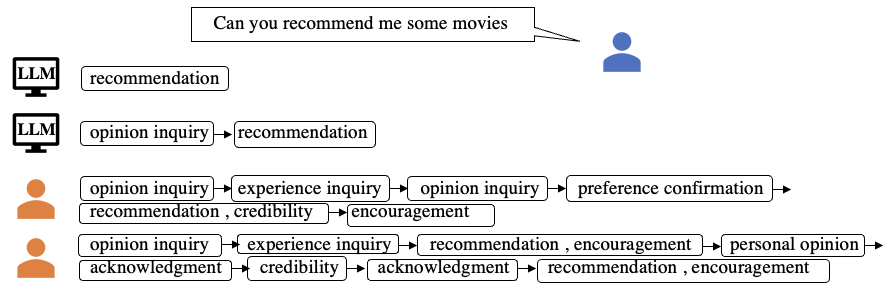}
    \caption{Comparing the strategies that LLM typically uses and human recommenders use for making conversational recommendations.}
    \label{fig:human}
    \vspace{-4mm}
\end{figure}

To validate these findings, we conducted a comparative analysis involving two popular LLMs and a human recommender across 20 randomly selected datapoints from INSPIRED dataset~\cite{hayati2020inspired}, the only popular conversational recommendation dataset providing behavior labels. This comparison revealed a noticeable difference in the behavior of LLM-based CRS and human recommenders. Specifically, LLM-based CRS systems tend to be passive and inflexible, often rushing to make recommendations without conducting any inquiry. 
In contrast, human recommenders display much greater patience, dynamism, and adaptability. They show a wider range of complex information-seeking strategies contributing to recommendations, as illustrated in Figure~\ref{fig:human}. A more tangible illustration is that human recommenders averagely conduct 2.5 conversational turns before making their first recommendation, a number far exceeds those by GPT3.5\cite{ouyang2022training} or Llama2\cite{touvron2023llama}, as shown in Table~\ref{tab:strategy}. The user information obtained from initial inquiries helps human recommenders make better recommendations and achieve higher success rates.




\begin{table}[]
\begin{tabular}{lcc}
\hline
\multicolumn{1}{l|}{}       & \multicolumn{1}{l|}{\#Turns before Rec} & \multicolumn{1}{l}{Success Rate}  \\ \hline
\multicolumn{3}{l}{\textbf{LLM-based CRS}}                                                                                                                \\ \hline
\multicolumn{1}{l|}{GPT 3.5} & \multicolumn{1}{c|}{1.158}           & \multicolumn{1}{c}{15.8\%}                               \\
\multicolumn{1}{l|}{Llama 2} & \multicolumn{1}{c|}{1.000}           & \multicolumn{1}{c}{5.3\%}                                \\ \hline
\multicolumn{3}{l}{\textbf{Reference}}                                                                                                           \\ \hline
\multicolumn{1}{l|}{Human}  & \multicolumn{1}{c|}{2.500}           & \multicolumn{1}{c}{57.1\%}                                      \\ \hline
\end{tabular}
\caption{Comparing the behavior of LLM-based CRSs and human's behavior including: average length of conversational turns before making the first recommendation(\#Turns before Rec), average success rate.}
\label{tab:strategy}
\vspace{-9mm}
\end{table}




Recently, many studies on LLMs have focused on improving their behavior alignment with humans~\cite{ouyang2022training, kopf2023openassistant, bao2023tallrec}. The alignment between the generated sentences from CRS and human recommenders is also desirable for LLM-based CRS. First, better behavior alignment can enhance the user experience by creating a more efficient and proactively interactive conversation. More importantly, the ability to use complex recommendation strategies can help the system receive more user preference information and better understand user profiles, leading to more accurate recommendations~\cite{hayati2020inspired}. Despite the importance of measuring behavior alignment, the evaluation metrics currently used in CRS have no ability to measure this. 


To fill this gap, we propose a new evaluation metric: {\em Behavior Alignment}, a metric that {\em explicitly} evaluates how well LLM-based CRS's recommendation strategies are aligned with humans.  We also conduct experiments to demonstrate the effectiveness of the measure. In particular, Behavior Alignment has high agreement with the human preferences, and can differentiate the performance of different LLM-based CRS systems much better than the existing metrics. However,
one limitation of Behavior Alignment is the requirement of having human annotations of recommendation strategies, which can be costly and time-consuming to obtain.  To overcome this limitation, we propose a classification-based method
to {\em implicitly} estimate the behavior alignment without the annotations of recommendation strategies. Experiment results confirm the robustness of the proposed implicit estimation of the Behavior Alignment measure across multiple CRS datasets.

\section{Related Work}


Existing research practices primarily evaluate CRS's performance in two aspects: recommendation accuracy, and the quality of sentence generation. In terms of recommendation accuracy, most CRS studies adopt metrics from traditional recommender systems, like hit rates, precision, and recall, from IR community~\cite{zhou2021crslab, gao2021advances}.


To evaluate the quality of generated sentences, automatic metrics borrowed from dialogue systems, such as Perplexity, BLEU, and DIST, are used~\cite{chen2019towards, zhou2020towards, zhou2021crslab, wang2022towards}. However, their alignment with \textit{quality} and \textit{user satisfaction} in the context of CRS is questioned~\cite{jannach2023evaluating, zhang2020evaluating}. 

Under such questions, several manual evaluation metrics are proposed. Chen et al.~\cite{chen2019towards} advocated for \textit{Consistency}. Zhou et al.~\cite{zhou2020improving} proposed \textit{fluency} and \textit{informativeness} as substitutes for BLEU. However, with the introduction of LLM for building CRS~\cite{he2023large, li2023large, zhang2023user, friedman2023leveraging}, fluency and informativeness have lost their relevance as LLM can generate highly fluent sentences and the informativeness can be conveniently adjusted by prompting. In addition, manual evaluation’s inherent subjectivity and cost can pose a challenge. Consequently, recent works in LLM-based CRS~\cite{he2023large, li2023large} opted to only evaluate CRS based on accuracy metrics, and overlooked the quality of generated sentences.

\vspace{-2mm}
\section{Behavior Alignment: the Metric}

Users tend to find the conversations more natural and engaging with human recommenders. And those conversations often lead to better recommendations compared with automatic conversational recommender systems~\cite{hayati2020inspired}. This is largely because human recommenders usually deploy various sociable and preference elicitation strategies to grab more personal information of users~\cite{hayati2020inspired}. In addition, we and existing studies~\cite{liao2023proactive, zhang2023user} consistently observed a substantial disparity between human's strategies and those employed by LLM-based CRS where latter often appears naive and passive. Therefore, human recommenders' behaviors could become valuable references for the LLM-based CRS.

\vspace{-2mm}
\subsection{Definition}



Based on the assumption that human recommenders' behaviors can perform as good references for CRS, given a context (e.g., user's past interactions with a CRS), a CRS generate a response $C$.  And a human recommender writes
a response $H$ for the same context. Assuming we know the recommendation strategy of $C$ is $R_C$ and the recommendation strategy of $H$ is $R_H$, 
the behavior alignment score of the pair (i.e., $BA(C,H)$) can be computed as in Equation (1). In a test collection with $N$ generated responses, the 
system's Behavior Alignment score can be calculated using Equation (2).
\vspace{-2mm}

\begin{equation}
BA(C,H)=\begin{cases}

1\;\;if\;R_C=R_H
 \\
0\;\;if\;R_C \ne R_H
\end{cases}
\end{equation}

\begin{equation}
    Behavior\;Alignment = \frac{\sum_{k=2..N} BA(C_k,H_k) }{N}
\end{equation}

In Equation (2), the first interactive turn is not counted into Behavior Alignment since the start point of every conversation can be random in practice. After the first turn is finished, the behaviors of following turns should conditioned on the existing context.


 For recommendation strategy $R$, existing CRS studies~\cite{hayati2020inspired} have provided a comprehensive categorization that includes 13 mutually exclusive types. These types are: \textit{acknowledgment}, \textit{redibility}, \textit{encouragement}, \textit{experience inquiry}, \textit{offer help}, \textit{opinion inquiry}, \textit{personal experience}, \textit{personal opinion}, \textit{preference confirmation}, \textit{rephrase preference}, \textit{self modeling}, \textit{similarity}, and \textit{transparency}.

\section{Behavior Alignment: Effectiveness as An Evaluation Metric}

Behavior Alignment can measure the behavior differences between generated sentences from CRS and human recommenders, but it remains unclear whether such a metric is more desirable than existing metrics. To answer this question, we design two sets of experiments to evaluate (1) \textit{whether Behavior Alignment reflects user preferences};  and (2) \textit{whether it can differentiate well-performed systems from those poorly performed ones}.

\vspace{-1mm}
\subsection{Agreement with Human Preferences}
The first set of experiments aims to assess whether the evaluation results of Behavior Alignment are closely consistent with human preferences. Specifically, responses from two CRS systems are evaluated by Behavior Alignment and humans. A strong agreement between them would reflect the effectiveness of a metric. 

We randomly sampled 1,000 instances from INSPIRED dataset. The two CRS systems are built on two LLMs respectively: Falcon-7B~\cite{penedo2023falcon} and Llama2-7B~\cite{touvron2023llama}. They were specifically chosen due to the distinct training data they utilize, which leads to varying response styles in reaction to dialog histories. To gauge human preferences, we enlisted two annotators who have experience in conversational recommendations. They assumed the role of CRS users and annotated preferences on the responses generated by the two LLM-based CRS systems (i.e., same, A is better than B, or B is better than A). In addition to preferences between the two systems, we ask an expert trained with linguistic backgrounds to annotate the recommendation strategies for each generated response, which is necessary for computing Behavior Alignment. 


An ideal evaluation metric should yield ratings that highly correlate to human preferences~\cite{deriu2021survey}. We use Cohen's Kappa to measure such correlation. 
In addition to Cohen's Kappa, we employed the Bootstrap method to establish the 2.5\% to 97.5\% confidence interval.

\begin{figure}
    \centering
        \includegraphics[scale=0.27]{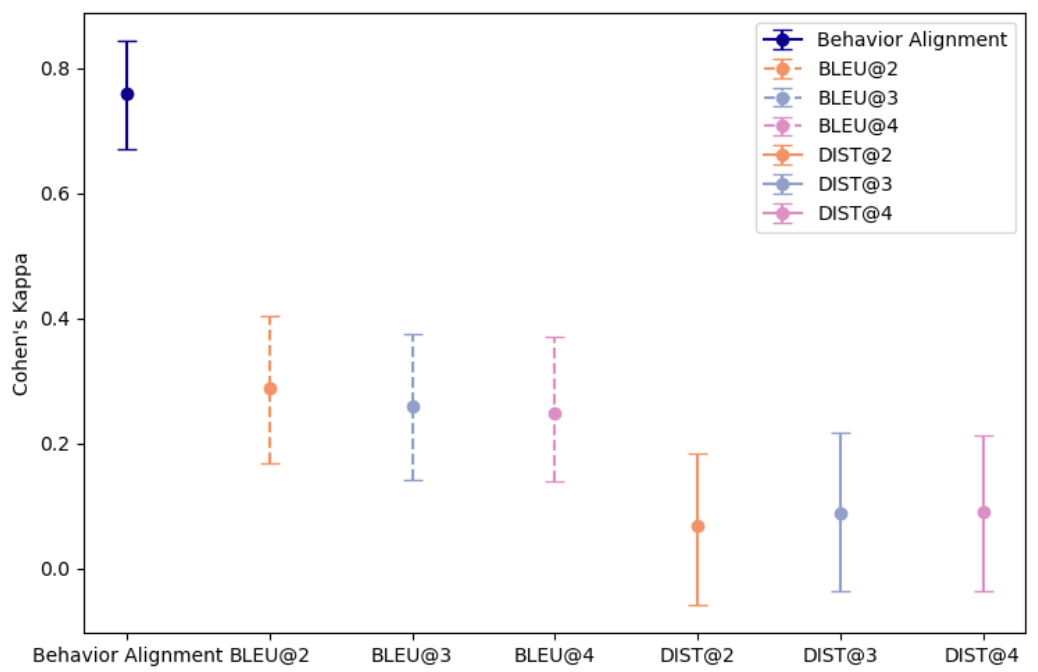}
        \caption{Agreement with human judgments over "Is the response from Falcon-based CRS better than Llama2-based CRS ?", measured by Cohen's Kappa.}
        \label{fig:cohen1}
                \vspace{-5mm}
\end{figure}

To compare how various metrics reveal user preferences, we conduct a comparative analysis with two popular automatic metrics that existing CRS studies used to evaluate \textit{generation quality}~\cite{chen2019towards, zhou2020improving, zhou2021crslab}: BLEU@K and DIST@K. The results are depicted in Figure~\ref{fig:cohen1}.


We can see Behavior Alignment demonstrates a considerable level of agreement with human preference, as evidenced by a Cohen's Kappa of 0.74, a value typically interpreted as indicating "Substantial Agreement". On the contrary, DIST@K and BLEU@K exhibit a much lower level of agreement with human preferences. One potential reason for the poor alignment is they are word-level metrics that are directly adopted from Machine Translation and Summarization tasks, which do not fit the context and goal of the conversational recommendation task very well~\cite{deriu2021survey}.

\begin{figure}
    \centering
        \includegraphics[scale=0.27]{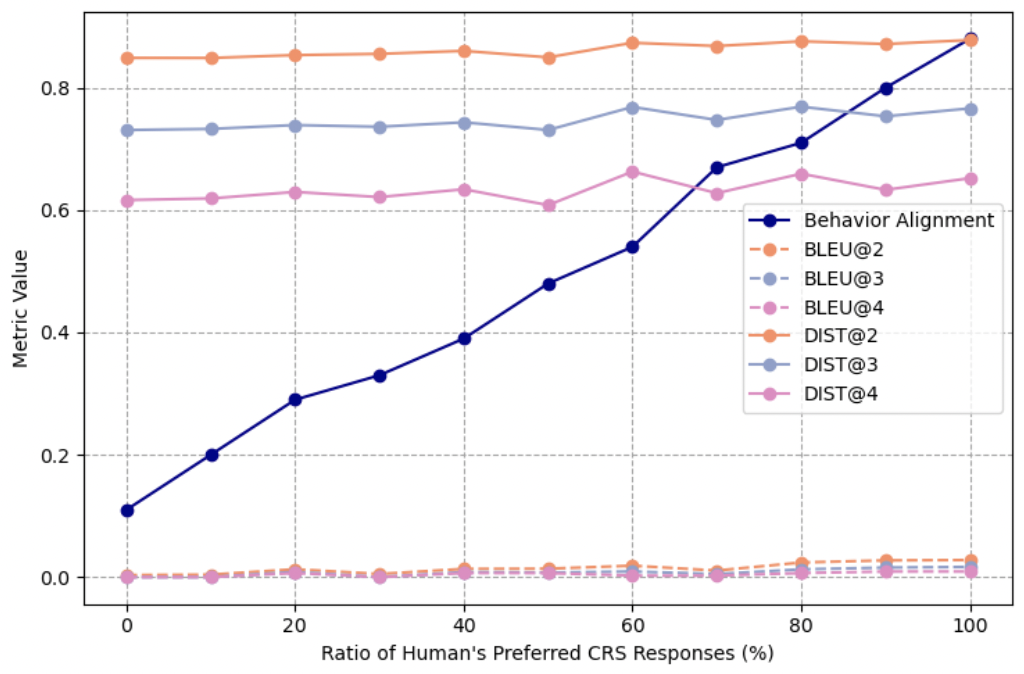}
        \caption{Metric score changes (y-axis) along with human preference changes (x-axis)}
        \label{fig:different}
                \vspace{-3mm}
\end{figure}

\vspace{-2mm}
\subsection{Differentiation of System Performance}

A desirable evaluation metric should be able to differentiate various systems~\cite{deriu2021survey}. 
Also, the increase/decrease of rating should be consistent with the increase/decrease of human preference for different systems. However, it is hard to quantify human preference for each LLM-based CRS and it is almost impossible to find large numbers of LLMs covering different levels of human preferences to meet the necessity for conducting evaluations.


\begin{table*}[]
\centering
\begin{tabular}{c|c|cc}
\hline
Behavior Type           & Accuracy & 1st Misclassification   & 2nd Misclassification \\ \hline
personal experience & 0.60     & credibility             & similarity            \\
rephrase preference  & 0.45     & preference confirmation & personal opinion      \\
self modeling       & 0.31     & personal experience     & similarity            \\
similarity          & 0.53     & acknowledgment          & self modeling         \\
transparency        & 0.65     & opinion inquiry         & offer help            \\ \hline
\end{tabular}
\caption{Top 1st and 2nd misclassification for those Behavior classes having accuracy lower than 0.7}
\label{table:misclass}
        \vspace{-7mm}
\end{table*}


To address these challenges, we build synthetic LLM-based CRS systems with different levels of user preferences by mixing different ratios of "good generations" and "bad generations" based on the annotated sentences generated from two LLM-based CRS we implemented. Specifically, 100 generation pairs were randomly chosen from annotated data, with one system's response identified as superior to the other. We selected the responses preferred by humans to create a hypothetical "ideal" system, while the responses that were less preferred were selected to build a "worst-case" system. Following this, we crafted synthetic CRS systems of varying quality by blending 90\%, 80\%, ..., 10\% of the "ideal" system with 10\%, 20\%, ..., 90\% of the "worst-case" system. After the synthesis, we compute each metric including Behavior Alignment to see if they can effectively differentiate between different synthetic systems.

As displayed in Figure~\ref{fig:different}, the x-axis signifies the proportion of samples taken from the "ideal" system, which can be interpreted as a "human preference score". The results show that there is a consistent increase in Behavior Alignment which ranges from 0.11 to 0.88 as we shift from a "worst-case" system to an "ideal" system. This rise is directly proportional to each incremental inclusion of human-preferred samples. However, the other two metrics, BLEU@K and DIST@K, exhibited minimal variability despite the enhancement in human-preferred samples. 

\section{Estimation of Implicit Behavior Alignment}

Although the proposed Behavior Alignment is effective, it requires 
real-time annotations of behavior types, which is not readily available.
To address this issue, we propose a classification-based method to 
estimate the implicit behavior alignment. Instead of directly predicting the behavior categories and then comparing whether
the behaviors of two responses are the same, the method focuses on a binary problem: predicting whether the two responses belong to the same category, which would not require the explicit behavior labels anymore\footnote{The source code and the experiments can be found at \url{https://github.com/dayuyang1999/Behavior-Alignment}}.

\vspace{-2mm}
\subsection{Methodology}

Given a human-generated response $H$ and a CRS-generated response $C$, we propose estimating the alignment through a binary classifier, which takes 
$H$ and $C$ as input and predicts whether they correspond to the same category or not. The binary classifier is built based on BERT\footnote{https://huggingface.co/bert-large-uncased} and fine-tuned on 100,000 human response pairs selected from INSPIRED dataset.  How to select the pairs?  We explore two strategies to construct the training/testing data.  

The first training strategy is straightforward. We created 50,000 negative sentence pairs with different behavior types and another 50,000 positive pairs with the same behavior type. For each sentence pair, they are concatenated by a special token [SEP], which allows BERT to identify the segment of two sentences. This strategy is referred to as "Original". 


The second strategy is to intentionally include more difficult examples in the training data to improve the robustness of the classifier.  
To achieve this, we first train a multi-class classifier to predict the exact behavior type of each sentence. The prediction results are compared with the ground truth labels in order to compute the misclassification rate. Specifically, we define "hard negatives" as negative pairs that a sentence having a prediction accuracy of less than 0.7 and another sentence is from its most misclassified category as shown in Table~\ref{table:misclass}.
Consequently, we created 10,000 hard negative samples across five categories. We then randomly replaced 10,000 negative examples in the original training data with hard negatives to build the "mixed-hard" training set.

\vspace{-2mm}
\subsection{Evaluation}

The robustness of a binary classifier is important, as it directly influences the efficacy and applicability of the proposed metric. To assess the robustness, we employ three methodologies: (1) implementing cross-validation to ensure consistently high accuracy; (2) determining its Cohen's Kappa in relation to human annotations to verify an invariably strong agreement; and (3) evaluating its performance on out-of-distribution data to ascertain minimal performance degradation.

\begin{table}[]
    \centering
        \begin{tabular}{l|cc}
        \hline
                 & \multicolumn{2}{l}{Accuracy on hold-out test set} \\ \cline{2-3} 
                 & Original               & Mixed-hard               \\ \hline
        Fold0    & 0.960                  & 0.976                    \\
        Fold1    & 0.958                  & 0.978                    \\
        Fold2    & 0.958                  & 0.977                    \\
        Fold3    & 0.940                  & 0.976                    \\
        Fold4    & 0.967                  & 0.973                    \\ \hline
        Averaged & 0.957                  & 0.976                    \\ \hline
        \end{tabular}
    \caption{Cross-validation results}
    \label{table:kfold}
    \vspace{-8mm}
\end{table}

First, we employ cross-validation on the binary sentence pairs. As observed in Table~\ref{table:kfold}, the accuracy remains consistently high regardless of which strategy is used for constructing training data. 

\begin{table}[]
    \centering
        \begin{tabular}{l|cc}
        \hline
                   & Accuracy & Cohen's kappa \\ \hline
        Original       & 0.957            & 0.913         \\
        Mixed-hard & 0.976            & 0.952         \\ \hline
        \end{tabular}  
        \caption{The agreement between predictions and human annotations on testing samples}
        \label{table:cohens}
        \vspace{-7mm}
\end{table}

\begin{table}[]
    \centering
        \begin{tabular}{l|cc}
        \hline
                      & Accuracy & Cohen's kappa \\ \hline
        Original      & 0.782    & 0.563         \\
        Mixed-hard & 0.932    & 0.865         \\ \hline
        \end{tabular}
        \caption{The agreement between predictions and human annotations on out-of-distribution instances from ReDial dataset}
        \label{table:redial}
        \vspace{-8mm}
\end{table}

Second, we utilized Cohen’s Kappa to quantify the agreement between the classifier's predictions and human annotations. 
As illustrated in Table~\ref{table:cohens}, our results depict that Cohen’s Kappa exceeds 0.9 on the testing set, denoting an almost impeccable alignment between the classifier's predictions and human annotations.

Lastly, to verify the generalizability of the binary classifier, we selected another widely-used CRS dataset that the model has not seen during training, ReDial~\cite{li2018towards} for testing. Compared with INSPIRED dataset, ReDial dataset emphasizes more on making recommendations during chi-chat, rendering its linguistic style more colloquial. The out-of-distribution samples that have differences on using of language may introduce challenges to the classifier.
Given the absence of behavior annotation in the ReDial dataset, we first instructed two annotators to annotate ReDial in a manner similar to INSPIRED, resulting in 42006 sentence pairs with ground-truth binary labels. As Table~\ref{table:redial} showcases, the binary classifier trained on "Original" sentence pairs witnessed a marked decline in performance. Conversely, the classifier trained on sentence pairs augmented with hard negatives demonstrated superior generalizability, achieving over 93\% accuracy and a Cohen’s kappa of 0.86 with humans. This generally indicates a near-perfect agreement between the automatic annotator and human recommenders. The findings emphasize the importance of hard negatives, particularly for a dataset-agnostic evaluation metric.

\vspace{-2mm}
\section{Conclusion}
Evaluating the generated sentences is crucial for CRS as they significantly influence user responses. This, in turn, impacts both information exposure and user profiling. In this paper, we introduce a novel evaluation metric designed to assess the quality of responses generated by LLM-based CRSs. Our experiments demonstrate this metric's effectiveness, particularly in its alignment with human preferences and its capacity to distinguish LLM-based CRSs by their ability to implement complex recommendation strategies.



\section{Acknowledgements}
This research is supported by the graduate fellowship from the Institute for Financial Services Analytics at the University of Delaware. We would also like to express our sincere gratitude to the reviewers for their insightful comments and suggestions.

\balance
\printbibliography

\end{document}